\documentclass{elsart}

\usepackage{graphicx}

\begin{document}

\begin{frontmatter}

\title{High Conductance States in a Mean Field Cortical Network Model}

\author{Alexander Lerchner}
\ead{lerchner@nordita.dk}
\author{Mandana Ahmadi}
\ead{ahmadi@nordita.dk}
\author{John Hertz}
\address{Nordita, Blegdamsvej 17, 2100 Copenhagen, Denmark}
\ead{hertz@nordita.dk}

\begin{abstract}
Measured responses from visual cortical neurons show that spike
times tend to be correlated rather than exactly Poisson
distributed. Fano factors vary and are usually greater than 1 due
to the tendency of spikes being clustered into bursts. We show
that this behavior emerges naturally in a balanced cortical
network model with random connectivity and conductance-based
synapses. We employ mean field theory with correctly colored noise
to describe temporal correlations in the neuronal activity. Our
results illuminate the connection between two independent
experimental findings: high conductance states of cortical neurons
in their natural environment, and variable non-Poissonian spike
statistics with Fano factors greater than 1.

\end{abstract}

\begin{keyword}
synaptic conductances \sep response variability \sep cortical
dynamics
\end{keyword}

\end{frontmatter}

\section{Introduction}

Neurons in primary visual cortex show a large increase in input
conductance during visual activation: \emph{in vivo} recordings
(see, e.g.,~\cite{Borg-Graham+Monier+Fregnac:1998}) show that the
conductance can rise to more than three times that of the resting
state. Such \emph{high conductance states} lead to faster neuronal
dynamics than would be expected from the value of the passive
membrane time constant, as pointed out by Shelley et
al.~\cite{Shelley+McLaughlin+Shapley+Wielaard:2002}. We use mean
field theory to study the firing statistics of a model network
with balanced excitation and inhibition and observe consistently
such high conductance states during stimulation.

In our study, we classify the irregularity of firing with the Fano
factor $F$, defined as the ratio of the variance of the spike
count to its mean. For temporally uncorrelated spike trains (i.e.,
Poisson processes) $F=1$, while $F>1$ indicates a tendency for
spike clustering (bursts), and $F<1$ points to more regular firing
with well separated spikes. Observed Fano factors for spike trains
of primary cortical neurons during stimulation are usually greater
than 1 and vary within an entire order of magnitude (see,
e.g.,~\cite{Gershon+Wiener+Letham+Richmond:1998}). We find the
same dynamics in our model and are able to pin-point the relevant
mechanisms: synaptic filtering leads to spike clustering in states
of high conductance (thus $F>1$), and Fano factors depend
sensitively on variations in both threshold and synaptic time
constants.

\section{The Model}

We investigate a cortical network model that exhibits
self-consistently balanced excitation and inhibition. The model
consists of two populations of neurons, an excitatory and an
inhibitory one, with dilute random connectivity. The model neurons
are governed by leaky integrate-and-fire subthreshold dynamics
with conductance-based synapses. The membrane potential of neuron
$i$ in population $a$ ($a = 1,2$ for excitatory and inhibitory,
respectively) obeys

\begin{equation}
    \label{eq:duFull}
    \frac{du_a^i(t)}{dt} = -g_L u_a^i(t) - \sum_{b=0}^2 \sum_{j=1}^{N_b}
    g_{ab}^{ij}(t)(u_a^i(t) - V_b).
\end{equation}

The first sum runs over all populations $b$, including the
excitatory input population representing input from the LGN and
indexed by $0$. The second sum runs over all neurons $j$ in
population $b$ of size $N_b$. The reversal potential $V_b$ for the
excitatory inputs ($b = 0,1$) is higher than the firing threshold,
the one for the inhibitory inputs ($V_2$) is below the reset
value. The constant leakage conductance $g_L$ is the inverse of
the membrane time constant $\tau_m$.

The time dependent conductance $g_{ab}^{ij}(t)$ from neuron $j$ in
population $b$ to neuron $i$ in population $a$ is taken as
\begin{equation}
    \label{eq:conductance}
    g_{ab}^{ij}(t) = \frac{g_{ab}^0}{\sqrt{K_b}} \sum_s
    \exp(-(t-t^j_s)/\tau_b) \Theta (t-t^j_s)
\end{equation}
if there is a connection between those two neurons, otherwise
zero. The sum runs over all spikes $s$ emitted by neuron $j$,
$\tau_b$ is the synaptic time constant for the synapse of type $b$
(excitatory or inhibitory), and $\Theta$ is the Heavyside step
function. $K_b$ denotes the average number of presynaptic neurons
in population $b$. We followed van Vreeswijk and
Sompolinsky~\cite{vanVreeswijk+Sompolinsky:1996} by scaling the
conductances with $1/\sqrt{K_b}$ so that their fluctuations are of
order one, independent of network size.

\section{Mean Field Theory}

We use mean field theory to reduce the full network problem to two
neurons: one for each population. This method is exact in the
limit of large populations with homogeneous connection
probabilities \cite{Mari:2000}. The neurons receive
self-consistent inputs from their cortical environment, exploiting
the fact that all neurons within a population exhibit the same
firing statistics due to homogeneity. The time dependent
conductance described in (\ref{eq:conductance}) can then be
replaced by a realization of a Gaussian distributed random
variable $g_{ab}$ with mean
\begin{equation}
    \label{eq:mean}
    \langle g_{ab} \rangle =  g_{ab}^0 \sqrt{K_b} \, r_b,
\end{equation}
and covariance
\begin{equation}
    \label{eq:covar}
    \langle \delta g_{ab}(t) \, \delta g_{ab}(t')\rangle =
    (g_{ab}^0)^2 (1-K_b/N_b) \, C_b(t-t'),
\end{equation}
Here, $r_b$ is the firing rate of the presynaptic neuron $b$, and
$C_b(t-t')$ is the autocorrelation function of its spike train. A
simple approximation of the autocorrelation, like the one used
by~\cite{Amit+Brunel:1997} and~\cite{Brunel: 2000}, is to assume
$g_{ab}(t)$ to be temporally uncorrelated (i.e., white noise), in
which case it simplifies to $C_b(t-t') = r_b \, \delta(t-t')$. The
term $(1-K_b/N_b)$ is a correction for the finite connection
concentration $K_b/N_b$ and can be derived using the methods
of~\cite{Kree+Zippelius:1987}.

The self-consistent balance condition is obtained by setting the
net current in~(\ref{eq:duFull}) to zero when the membrane
potential is at threshold $\theta_a$ and the conductances have
their mean values~(\ref{eq:mean}). In the large $K_b$-limit, it
reads
\begin{equation}
    \label{eq:balance}
    \sum_{b=0}^2 g_{ab}^0 \sqrt{K_b} \, r_b \, (\theta_a - V_b) = 0.
\end{equation}

The distribution of the variables $g_{ab}$ can be calculated
numerically using an iterative
approach~\cite{Eisfeller+Opper:1992}. One starts with a guess
based on the balance equation~(\ref{eq:balance}) for the means and
covariances and generates a large sample of specific realizations
of $g_{ab}(t)$, which are used to integrate (\ref{eq:duFull}) to
generate a large sample of spike trains. The latter can then be
used to calculate new estimates of the means and covariances by
applying (\ref{eq:mean}) and (\ref{eq:covar}) and correction of
the initial guess towards the new values. These steps are repeated
until convergence.

\section{Results}

For the above described model, we chose parameters corresponding
to population sizes of 16,000 excitatory neurons and 4,000
inhibitory neurons, representing a small patch of layer IV cat
visual cortex. The neurons were connected randomly, with 10\%
connection probability between any two neurons. The firing
threshold was fixed to 1, excitatory and inhibitory reversal
potentials were set to $+14/3$ and $-2/3$, respectively, and the
membrane time constant $\tau_m = g_L^{-1}$ was $10$~ms. For the
results presented here, the integration time step was 0.5 ms.

\begin{figure}[t]
 \includegraphics[width=13.6cm,height=9cm]{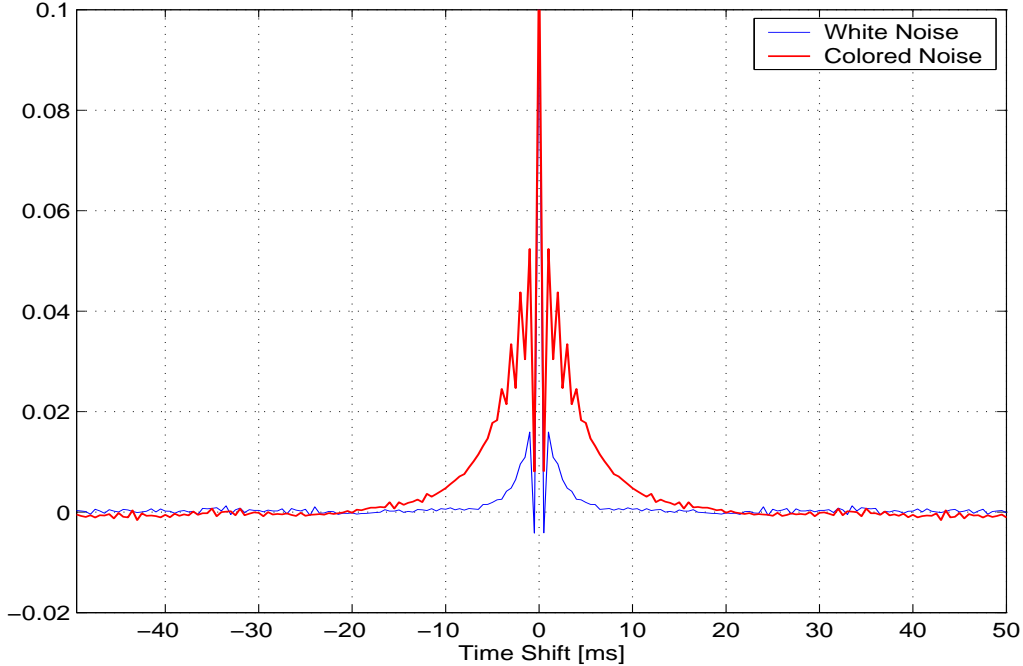}
\caption{Autocorrelation functions for white noise (blue) and
colored noise (red). The white noise approximation underestimates
the amount of temporal correlation in the neuron's firing.}
\label{fig:Autocorr}
\end{figure}

Figure~\ref{fig:Autocorr} illustrates the importance of coloring
the noise produced by intra-cortical activity. The white noise
approximation underestimates both the correlation times and the
strength of the correlations in the neuron's firing: its
autocorrelation (blue) is both narrower and weaker than the one
for colored noise (red).

\begin{figure}[t]
 \includegraphics[width=13.6cm,height=9cm]{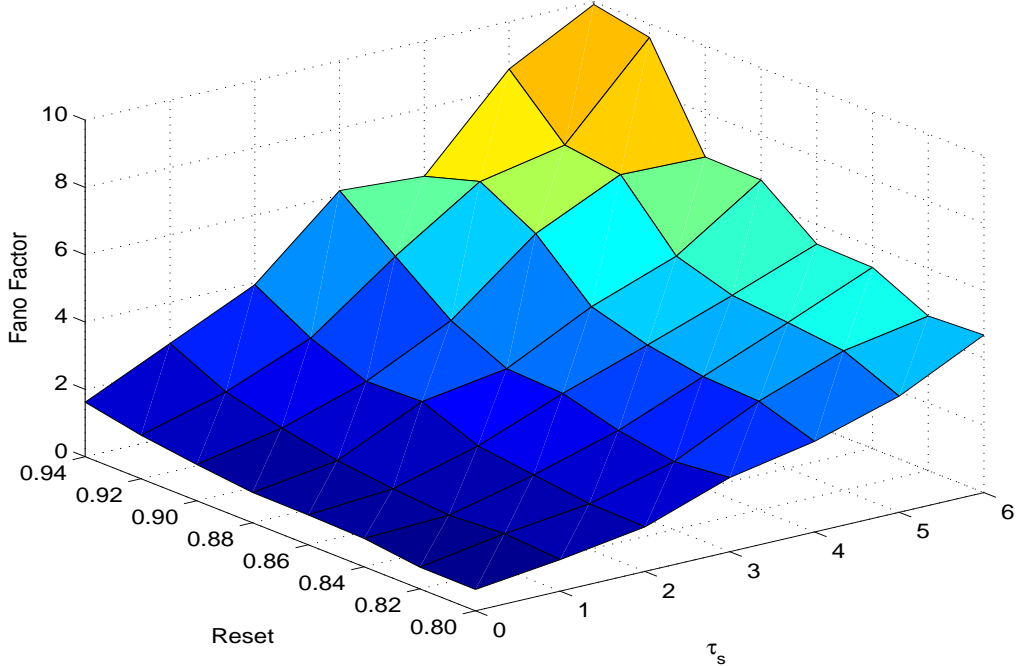}
\caption{Fano factors for a range of reset values and synaptic
time constants $\tau_s$. Longer synaptic time constants lead to
increased clustering (bursts) of spikes, which is reflected in
higher Fano factors.}
\label{fig:Fanos}
\end{figure}

Fano factors vary systematically with both the distance between
reset and threshold and the synaptic time constant $\tau_s$.
Non-zero synaptic time constants produced consistently Fano
factors greater than one. We varied the reset between 0.8 and 0.94
and $\tau_s$ between 0 and 6~ms, which resulted in values for $F$
that span an entire order of magnitude, from slightly above 1 to
approximately 10 for $\tau_s \ge 2$~ms (see
Figure~\ref{fig:Fanos}).

\section{Discussion}

\begin{figure}[t]
 \includegraphics[width=13.6cm,height=9cm]{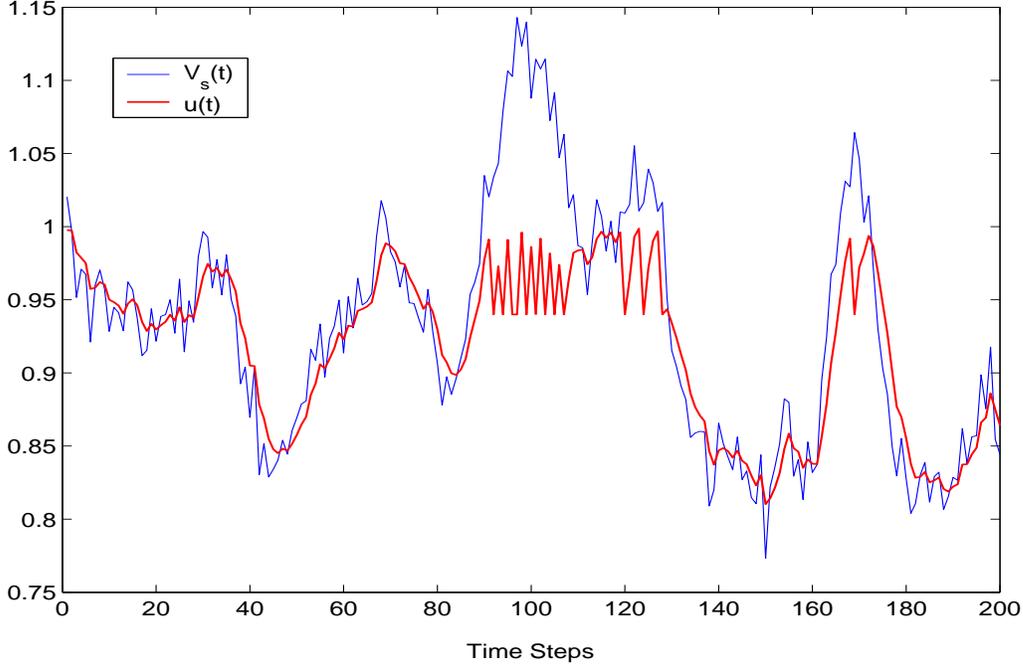}
\caption{The membrane potential $u(t)$ (red) follows the effective
reversal potential $V_S(t)$ (blue) closely, except for detours
when the neuron is reset due to firing. The membrane potential
recovers fast enough to spike several times while $V_S(t)$ stays
above threshold, thus producing bursts of spikes. Here, the
threshold is set to 1 and the reset to 0.94.}
\label{fig:VVs}
\end{figure}

In all our simulations, we observed that the membrane potential
changed on a considerably faster time scale than the membrane time
constant $\tau_m = 10$~ms. This behavior is only observed if
conductance-based synapses are included in the integrate-and-fire
neuron model. To understand this phenomenon, it is convenient to
follow the notation of Shelley et
al.~\cite{Shelley+McLaughlin+Shapley+Wielaard:2002} to rewrite the
equation for the membrane potential dynamics (\ref{eq:duFull}) in
the following form:
\begin{equation}
    \frac{du_a(t)}{dt} = -g_T(t) \left( u_a(t)-V_S(t) \right),
    \label{eq:duShort}
\end{equation}
with the \emph{total conductance} $g_T(t) = g_L + \sum_b
g_{ab}(t)$, and the \emph{effective reversal potential} $V_S(t) =
g_T(t)^{-1} \sum_b g_{ab}(t)V_b$. The membrane potential $u_a(t)$
follows the effective reversal potential with the input dependent
\emph{effective membrane time constant} $g_T(t)^{-1}$. The
effective reversal potential changes on the time scale of the
synaptic time constants, which are up to five times shorter than
$\tau_m$ in our simulations. However, if the effective membrane
time constant is shorter than the synaptic time constant due to a
large enough total conductance, then $u_a(t)$ can follow $V_S(t)$
closely, as observed in our simulations (see
Figure~\ref{fig:VVs}).

In high conductance states, the firing statistics are strongly
influenced by synaptic dynamics (see Figure~\ref{fig:Fanos}). This
is in contrast with strictly current based models, where the
neuron reacts too slow to reflect fast synaptic dynamics in its
firing. The `synaptic filtering' of arriving spikes leads to
temporal correlations in $V_S(t)$ and thus to temporal
correlations (by way of spike clustering) in firing. Therefore,
the model neurons receive temporally correlated input rather than
white noise. For this reason, in mean field models dealing with
conductance based dynamics, coloring the noise is important to
arrive at the full amount of temporal correlation in firing
statistics (see Figure~\ref{fig:Autocorr}). We confirmed these
considerations by running simulations without synaptic filtering
($\tau_s=0$). As expected, intra-cortical activity became
uncorrelated and the white noise approximation produced the same
result as coloring the noise correctly. In that case, Fano factors
stayed close to 1 (see Figure~\ref{fig:Fanos}), i.e, no tendency
of spike clustering was observed.

Previous investigations showed that varying the distance between
threshold and reset in balanced integrate-and-fire networks has a
strong effect on the irregularity of the
firing~\cite{Hertz+Richmond+Nilsen:2002}. By including a
conductance-based description of synapses, we were now able to
show the importance of synaptic time constants on firing
statistics, even if they are several times smaller than the
passive membrane time constant: Synaptic filtering facilitates
clustering of spikes in states of high conductance.

\end{document}